\documentclass[11pt,preprintnumbers,aps,amssymb,nofootinbib,amsmath,superscriptaddress,notitlepage,prd]
{revtex4-1}
\usepackage{epsfig,epsf}
\usepackage{bm} 
\usepackage{color} 
\definecolor{darkgreen}{RGB}{0,128,0}
\definecolor{darkred}{RGB}{139,0,0}
\usepackage{slashed}
\usepackage{hyperref}
\newcommand{\beq}{\begin{equation}}
\newcommand{\beql}[1]{\begin{equation}\label{#1}}
\newcommand{\eeq}{\end{equation}}
\def\bal#1\gal{\begin{align}#1\end{align}}
\newcommand{\ball}[1]{\bal\label{#1}}
\newcommand{\eq}[1]{(\ref{#1})}
\newcommand{\fig}[1]{Fig.~\ref{#1}}
\renewcommand{\sec}[1]{Sec.~\ref{#1}}
\renewcommand{\b}[1]{{\bm #1}} 
\newcommand{\unit}[1]{\hat {{\bm #1}}} 

\begin{document}


\title{The impact of  domain walls on the chiral magnetic effect  in hot QCD matter}

\author{Kirill Tuchin}

\affiliation{Department of Physics and Astronomy, Iowa State University, Ames, IA 50011, USA}

\date{\today}

\pacs{}

\begin{abstract}

The Chiral Magnetic Effect (CME) -- the separation of positive and negative electric charges along the direction of the external magnetic field in quark-gluon plasma and other topologically non-trivial media  -- is a consequence of the coupling of electrodynamics to the topological gluon field fluctuations that form metastable $CP$-odd domains. In phenomenological models it is usually assumed that the domains are uniform and the influence of the domain walls on the electric current flow is not essential. This paper challenges the latter assumption. A simple model consisting of a uniform spherical domain in a uniform time-dependent magnetic field is introduced and analytically solved. It is shown that (i) no electric current flows into or out of the domain, (ii)  the charge separation current, viz.\ the total electric current flowing inside the domain in the external field direction,  is a dissipative Ohm current, (iii) the CME effect can be produced either by the anomalous current or by the boundary conditions on the domain wall and 
(iv) the charge separation current oscillates in plasma long after the external field decays. These properties are qualitatively different from the CME in an infinite medium.

\end{abstract}

\maketitle

\section{Introduction}\label{sec:i}

The chiral magnetic effect (CME) is induction of electric current along the direction of the applied magnetic field \cite{Kharzeev:2004ey,Kharzeev:2007jp,Kharzeev:2007tn,Fukushima:2008xe,Kharzeev:2009fn}. It occurs in topologically non-trivial systems with chiral anomaly \cite{Adler:1969gk,Bell:1969ts} and breaks local $P$ and $CP$ symmetries. A phenomenological manifestation of CME is separation of positive and negative electric charges along the magnetic field direction \cite{Kharzeev:2007jp}. In relativistic heavy-ion collisions, electric charges  in quark-gluon plasma (QGP) separate along the direction of the external magnetic field created by the spectator valence quarks \cite{Kharzeev:2007jp,Skokov:2009qp,Voronyuk:2011jd,Ou:2011fm,Bzdak:2011yy,Bloczynski:2012en,Deng:2012pc,Tuchin:2010vs,Tuchin:2015oka,Stewart:2017zsu,Peroutka:2017esw}. There are several phenomenological approaches that link this effect to the experimental data \cite{Hirono:2014oda,Yin:2015fca}. 

Quantitative analysis of the charge separation requires knowledge of the medium response to the external electromagnetic field. The simplest model is to add a new anomalous current $\b j_A= \sigma_\chi \b B$ to the Amper law, where the chiral conductivity $\sigma_\chi$ is assumed to be weakly dependent on position and time \cite{Fukushima:2008xe,Kharzeev:2009fn,Kharzeev:2009pj}. The time dependence of the chiral conductivity arises primarily due to  the sphaleron transitions, finite quark mass and the helicity exchange between the magnetic field and QGP.  All these effects have very long characteristic time scales compared to the QGP lifetime \cite{Hirono:2015rla,Tuchin:2014iua,Bodeker:1998hm,Arnold:1998cy,Grabowska:2014efa}, which justifies treating $\sigma_\chi$ as time-independent.\footnote{Other, more exotic, effects that may induce time-dependence are discussed in \cite{Hirono:2016jps,Tuchin:2016tks}. } The assumption of the spatial uniformity is less sound however. The topological  $CP$-odd fluctuations of the hot nuclear matter occupy a region of a typical size $\sim 1/g^2T$ which is of the order of a fm. This implies that a typical heavy-ion collision can produce a large number of topologically different metastable $CP$-odd domains. Electric current varies steeply between the domain interior and the surrounding plasma. Thus, the charge separation effect is expected to be strongly dependent on the domain size and topology. The main goal of this paper is to compute the charge separation current taking into account these finite size effects. 

In order to study the charge separation effect in a finite size domain, it is advantageous to consider an exactly solvable model. The model considered in this paper consists of a spatially uniform spherical domain  of radius $R$ immersed into a topologically trivial environment. The electrodynamics with the chiral anomaly is described by the Maxwell-Chern-Simons theory (MCS)   in which the anomalous terms are associated with the background pseudoscalar field $\Theta$ whose dynamical extension is the axion \cite{Wilczek:1987mv,Carroll:1989vb, Sikivie:1984yz,Kharzeev:2009fn}. The role of the chiral anomaly is twofold:  it induces a new anomalous current   into the Amper law and causes a discontinuity of the normal electric and tangential magnetic field components at the domain wall even in the absence of the surface currents. Thus, the computation of the charge separation current entails solving the MCS equations inside the domain, in the presence of the anomalous current, and outside the domain and matching these solutions by means of the boundary conditions.

The paper is structured is follows.  The basic equations of the MCS theory and the corresponding boundary conditions are discussed in \sec{sec:a}. Considering a spatially uniform domain of an arbitrary shape, it is shown that the boundary conditions require vanishing of the normal component of the current on the domain wall.  General solutions to the MCS equations inside and outside a domain are obtained in \sec{sec:b} for a uniform monochromatic external field. Then in \sec{sec:c} these solutions are matched using the boundary conditions which yield analytical expressions for the magnetic field spectrum in entire space.  The result of \sec{sec:c} allows one to compute the induced magnetic field for any time-dependence of the external magnetic field. The analytical expressions for the total electric current   flowing through any cross section of the domain perpendicular to the external field direction \eq{d10} and the magnetic moment of the domain are also derived.  This is used in \sec{sec:e} to numerically compute the magnetic field of the domain using the known  time-dependence of the external magnetic field produced in relativistic heavy-ion collisions.   The results are summarized and discussed  in \sec{sec:s}.

\section{Field equations and boundary conditions}\label{sec:a}

The field equations of electrodynamics coupled to the topological charge carried by the gluon field read \cite{Wilczek:1987mv,Carroll:1989vb, Sikivie:1984yz,Kharzeev:2009fn} 
\begin{subequations}\label{a2}
\bal
&\b \nabla\cdot \b B=0\,, \label{a3}\\
& \b \nabla\cdot (\b E+c_A\Theta \b B)= 0\,,  \label{a4}\\
& \b \nabla \times \b E= -\partial_t \b B\,,\label{a5}\\
& \b \nabla \times (\b B-c_A\Theta \b E)= \partial_t (\b E+c_A\Theta \b B)+ \b j \,,\label{a6}
\gal
\end{subequations}
where $c_A=N_c\sum_f q_f^2 e^2/2\pi^2$ is the chiral anomaly coefficient. The plasma is assumed to be electrically neutral. The Ohm current  is $\b j= \sigma \b E$ where $\sigma$ is the electrical conductivity. The background field $\Theta$ is regarded as spatially uniform everywhere except the domain wall where  $\b \nabla \Theta$ is discontinuous.  

As explained in Introduction, the time-variation of $\Theta$ is too slow to be important for the heavy-ion phenomenology.  Nevertheless, since the chiral conductivity is proportional to the time-derivative of $\Theta$ one needs to keep track of its small variations. Hence $\Theta$ is approximated by \cite{Tuchin:2014iua}
\ball{a10}    
\Theta\approx \Theta_0+\mu_5 t\,,
\gal
where $\mu_5$ is the axial chemical potential related to the chiral conductivity $\sigma_\chi$ as $\mu_5=\sigma_\chi/c_A$ \cite{Fukushima:2008xe,Kharzeev:2009fn}. Estimating the chiral conductivity  optimistically as  $\sigma_\chi=10^{-2}\,\mathrm{fm}^{-1}$ and using $c_A=1/129$ one obtains $\mu_5=1.3\,\mathrm{fm}^{-1}$. Thus, the time-dependent term in \eq{a10} is smaller than $2\pi$ for $t<3$~fm.  From now on it is assumed that this condition is satisfied. 

With the assumptions outlined in the preceding paragraphs one can simplify equations \eq{a3}-\eq{a6}, which read at any point in space except the domain wall
\begin{subequations}\label{a11}
\bal
&\b \nabla\cdot \b B=0\,, \label{a12}\\
& \b \nabla\cdot \b E= 0\,,  \label{a13}\\
& \b \nabla \times \b E= -\partial_t \b B\,,\label{a15}\\
& \b \nabla \times \b B= \partial_t \b E+ \sigma_\chi \b B+\b j \,.\label{a16}
\gal
\end{subequations}
The assumption of the uniformity of the domain interior means that its wall width is neglected. The boundary conditions on the domain wall can be obtained directly  from equations \eq{a3}-\eq{a6}. Denoting by $\Delta$ the discontinuity of a field component across the domain wall and  neglecting the time-dependent term in \eq{a10}  one obtains \cite{Sikivie:1984yz}
\begin{subequations}\label{a21}
\bal
&\Delta B_\bot=0\,, \label{a22}\\
& \Delta  (E_\bot+c_A\Theta_0 B_\bot)= 0\,,  \label{a23}\\
&  \Delta \b E_\parallel= 0\,,\label{a24}\\
& \Delta (\b B_\parallel-c_A\Theta_0 \b E_\parallel)= 0 \,.\label{a25}
\gal
\end{subequations}
where $E_\bot$, $B_\bot$  and $\b E_\parallel$, $\b B_\parallel$ are components of the electromagnetic field normal and tangential to the domain wall respectively.  

A more stringent boundary condition can be derived using the continuity equation $\b \nabla\cdot \b j=0$, which implies that $\Delta j_\bot=0$ \cite{CK}. Projecting \eq{a6} onto the normal direction and using \eq{a5} one obtains
\ball{a30}
(\b \nabla\times \b B)_\bot+c_A\Theta\partial_t  B_\bot-c_A(\b\nabla\Theta \times \b E)_\bot= \partial_t(E+c_A\Theta B)_\bot +j_\bot\,.
\gal
The third term on the left-hand side vanishes because $\b \nabla\Theta$ points in the normal direction. The terms on the right-hand side are continuous in view of \eq{a23}. Now, solutions of \eq{a16} is a complete set of eigenstates of the curl operator satisfying the equation $\b\nabla\times \b B= \alpha \b B$, where $\alpha$  depends on medium properties. Consider such an eigenstate of frequency $\omega$. Then \eq{a30} implies that $B_\bot(\alpha+i\omega c_A\Theta)$  is continuous across the wall. However, $B_\bot$ is also continuous, whereas $\alpha$ and $\Theta$ are discontinuous. These conditions can only be satisfied  if $ B_\bot$ vanishes on the wall:
\ball{a33}
 B_\bot\big|_\mathrm{wall}=0\,.
 \gal

\section{Electromagnetic field of a spherical domain in uniform monochromatic magnetic field}\label{sec:b0}

\subsection{General solution inside and outside domain}\label{sec:b}

The external homogeneous magnetic field of frequency $\omega$ induces electromagnetic field in the domain which is governed by equations \eq{a11} and boundary conditions \eq{a21},\eq{a33}. Since electric and magnetic fields are divergentless, it is convenient to use the radiation gauge $\b \nabla \cdot \b A=0$, $A^0=0$ which allows one to write \eq{a16} as an equation for the vector potential 
\ball{b1}
\nabla^2 \b A= \partial_t^2\b A+\sigma \partial_t\b A-\sigma_\chi \b \nabla\times \b A\,.
\gal
Separation of  the temporal dependence of the vector-potential  $\b A(\b x,t)= \b A_\omega(\b x) e^{-i\omega t}$ yields for its monochromatic component 
\ball{b2}
\nabla^2 \b A_\omega= -\omega(\omega+i\sigma) \b A_\omega-\sigma_\chi \b \nabla\times \b A_\omega\,.
\gal
The general solution of \eq{b2} can be written as a superposition of the eigenfunctions of the curl operator. These functions are denoted by   $\b W^{\pm}_{lm}(\b x,\alpha)$ and satisfy the equation
 \ball{b5}
\b\nabla\times \b W^\pm_{lm}(\b x,\alpha)= \pm \alpha  \b W^\pm_{lm}(\b x,\alpha)\,.
\gal
Their explicit form in the spherical coordinates reads \cite{CK}
\ball{b8}
\b W_{lm}^\pm (\b x, \alpha)= \b T_{lm}(\b x, \alpha)\mp i \b P_{lm}(\b x, \alpha)\,,
\gal
where 
\bal
\b T_{lm}(\b x, \alpha)&= \frac{f_l(\alpha r)}{\sqrt{l(l+1)}}\left\{ -\frac{m}{\sin\theta}Y_l^m(\theta,\phi)\unit \theta-i\partial_\theta Y_l^m(\theta,\phi)\unit \phi\right\}\,, \label{b10}\\
\b P_{lm}(\b x, \alpha)&=\frac{1}{\sqrt{l(l+1)}}\left\{-\frac{l(l+1)}{\alpha r}f_l(\alpha r)Y_l^m(\theta,\phi)\unit r
-\frac{1}{\alpha r}\partial_r[f_l(\alpha r) r]\partial_\theta Y_l^m(\theta,\phi) \unit \theta  \right.\nonumber \\
&\left. 
-\frac{im}{\alpha \sin\theta}f_l(\alpha r) Y_l^m(\theta,\phi)\unit \phi\right\}\,.\label{b11}
\gal
$f_l$ is a linear combination of the spherical Bessel functions $j_l$ and $n_l$. 
The $z$-axis is chosen in the direction of the external magnetic field which  is given by 
\ball{b12}
\b B^\text{ext}= B_0\unit z e^{-i\omega t}= B_0(\cos\theta \unit r -\sin\theta \unit \theta)e^{-i\omega t}\,.
\gal 
The corresponding vector potential is
\ball{b13}
\b A^\text{ext}= \frac{1}{2}B_0r\sin\theta \unit \phi e^{-i\omega t}\,.
\gal
The symmetry considerations imply that in a spherical domain the only nontrivial component of the induced field is proportional to the linear combination of the functions
\bal
\b W_{10}^\pm(\b x, \alpha)=& -\frac{i}{\sqrt{2}}f_1(\alpha r)\partial_\theta Y_1^0(\theta,\phi)\unit \phi
\pm \frac{i\sqrt{2}}{\alpha r}f_1(\alpha r)Y_1^0(\theta,\phi)\unit r \nonumber \\
&\pm \frac{i}{\sqrt{2}\alpha r}\partial_r[f_1(\alpha r) r]\partial_\theta Y_1^0(\theta,\phi)\unit \theta\,. \label{b19}
\gal

The general solution to \eq{b2} inside the domain reads
\ball{b21}
\b A_\omega^\text{in}(\b x)= \sum_{lm}\left[ g_{lm} \b W_{lm}^+(\b x, q_+)+h_{lm}\b W_{lm}^-(\b x, q_-) \right]\,,
\gal
where $q_\pm$ are the roots of the equations $-q_\pm^2=-\omega(\omega+i\sigma)\mp \sigma_\chi q_\pm$. Namely,\footnote{The other two roots give linearly dependent solutions. They can be obtained by replacing $q_\pm \to -q_{\mp}$ which corresponds to $T_{lm}\to (-1)^l T_{lm}$,  $P_{lm}\to (-1)^{l+1} P_{lm}$.} 
\ball{b23}
q_\pm =\pm \frac{\sigma_\chi}{2}+\sqrt{(\sigma_\chi/2)^2+\omega(\omega+i\sigma)}\,.
\gal
The boundary conditions at the origin require that $f_l(q_\pm r)= j_l(q_\pm r)$. In view of \eq{b5}, the magnetic field inside the domain is
\ball{b25}
\b B_\omega^\text{in}(\b x)= & \sum_{lm}\left[ g_{lm} q_+\b W_{lm}^+(\b x, q_+)-h_{lm}q_-\b W_{lm}^-(\b x, q_-) \right]\,.
\gal

The general solution to \eq{b1} outside the domain, where $\Theta=0$, reads
\ball{b31}
\b A_\omega^\text{out}(\b x)= \sum_{lm}\left[ c_{lm} \b W_{lm}^+(\b x, k)+d_{lm}\b W_{lm}^-(\b x, k) \right]\,,
\gal
where $k = \sqrt{\omega(\omega+i \sigma)}$ and $f_l(kr)= \cos\delta_l j_1(kr)-\sin\delta_l n_l(kr)$. 
The magnetic field outside the domain is 
\ball{b35}
\b B_\omega^\text{out}(\b x)= \sum_{lm}k\left[ c_{lm} \b W_{lm}^+(\b x, k)-d_{lm}\b W_{lm}^-(\b x, k) \right]\,.
\gal
Note that \eq{b31} and \eq{b35} do not include the external field.

\subsection{Matching the solutions on the domain wall }\label{sec:c}

The boundary conditions \eq{a21},\eq{a33}  on the  spherical domain wall of radius $R$ read, after replacing $\b E_\omega= i\omega \b A_\omega$:
\begin{subequations}\label{c11}
\bal
&B_{\omega r}^\text{in}\big|_{r=R}= B_{\omega r}^\text{out}\big|_{r=R}+B_0\cos\theta=0\,,\label{c12}\\
&A_{\omega r}^\text{in}\big|_{r=R}= A_{\omega r}^\text{out}\big|_{r=R}\,,\label{c13}\\
&A_{\omega \theta}^\text{in}\big|_{r=R}= A_{\omega \theta}^\text{out}\big|_{r=R}\,,\label{c14}\\
&A_{\omega \phi}^\text{in}\big|_{r=R}= A_{\omega \phi}^\text{out}\big|_{r=R}+\frac{1}{2}B_0R\sin\theta \,,\label{c15}\\
&(B_{\omega \theta}^\text{in}+i\omega c_A\Theta_0 A_{\omega \theta}^\text{in})\big|_{r=R}=B_{\omega \theta}^\text{out}\big|_{r=R}-B_0\sin\theta\,,\label{c16}\\
&(B_{\omega \phi}^\text{in}+i\omega c_A\Theta_0 A_{\omega \phi}^\text{in})\big|_{r=R}=B_{\omega \phi}^\text{out}\big|_{r=R}\,.\label{c17}
\gal
\end{subequations}
Since the external magnetic field can be written as $\b B^\text{ext}_\omega= -\sqrt{6\pi}B_0 \b P_{10}(\b x,0)$, the only non-trivial solution to \eq{c11} is for the partial amplitudes with $l=1$ and $m=0$.  It easy to verify, using \eq{b21},\eq{b25},\eq{b31},\eq{b35} that the boundary conditions \eq{c12} and \eq{c15} are identical. Also, vanishing of $B_{\omega r}^\text{in}$ on the wall, i.e.\ \eq{c12}, implies vanishing of $A_{\omega\phi}$ on the wall, which in turn indicates that \eq{c13} and \eq{c17} are identical.  
Thus, there are five equations to determine five unknown amplitudes $g_{10}$, $h_{10}$, $c_{10}$, $d_{10}$ and $\delta_1$. It is understood that  $\Theta\neq 0$ inside the domain for otherwise  some of the equations \eq{c11} become redundant. 

To write the solution of the boundary conditions \eq{c11} in a compact form denote $\partial_r[j_1(\alpha r  ) r]|_{r=R}\equiv [j_1(\alpha R)R]'$ and define three auxiliary functions
\begin{subequations}\label{c24}
\bal
W_1&=j_1(Rq_+)\,[j_1(R q_- ) R]'-j_1(Rq_-)\,[j_1(R q_+ ) R]'\,,\label{c25}\\
W_2&= j_1(Rq_+)\,[j_1(R q_- ) R]'\, q_++j_1(Rq_-)\,[j_1(R q_+ ) R]'\,q_-\,,\label{c26}\\
W_3&=j_1(Rk)\,[n_1(R k) R]'-n_1(R k)\,[j_1(R k ) R]'\,.\label{c27}
\gal
\end{subequations}
After tedious but straightforward algebraic manipulations one obtains
\bal
g_{10}&= \sqrt{\frac{2\pi}{3}}B_0R\frac{k^2}{\omega}
\frac{W_2-2(q_++q_-)j_1(Rq_+)\, j_1(Rq_-)}{j_1(Rq_+)(q_++q_-)
\left[\frac{ik^2}{\omega}W_1+c_A \Theta W_2\right]}\,,\label{c33}\\
h_{10}&= -\frac{ j_1(Rq_+)}{ j_1(Rq_-)}g_{10} 
\,. \label{c34}
\gal
Eq.~\eq{c34} follows directly from the boundary condition \eq{a33}, or equivalently, \eq{c12}.
Other amplitudes can be expressed in terms of $g_{10}$. Define two more auxiliary functions
\begin{subequations}\label{c38}
\bal
a&= g_{10}\, \frac{q_++q_-}{k}j_1(Rq_+)\,, \label{c39}\\
b&= -\frac{g_{10}}{j_1(Rq_-)}\left[ W_1+W_2\left(\frac{1}{k}+\frac{ic_A\Theta\omega}{k^2}\right)\right]\,.\label{c40}
\gal
\end{subequations}
The amplitudes of the positive helicity component of the magnetic field outside the domain, see \eq{b35}, are
\begin{subequations}\label{c50}
\bal
c_{10}\cos\delta_1 &= -\frac{1}{2W_3}\left\{ \left(b+iRB_0 2\sqrt{2\pi/3}\right)n_1(Rk)-\left(a+iRB_0\sqrt{2\pi/3}\right)[n_1(Rk)R]'\right\}\,, \label{c51}\\
-c_{10}\sin\delta_1&= \frac{1}{2W_3}\left\{ \left(b+iRB_0 2\sqrt{2\pi/3}\right)j_1(Rk)-\left(a+iRB_0\sqrt{2\pi/3}\right)[j_1(Rk)R]'\right\}\,, \label{c52}
\gal
The ratio of these equations immediately yields $\tan\delta_1$. The remaining amplitudes, corresponding to the negative helicity component of the magnetic field outside the domain,  read
\bal
d_{10}\cos\delta_1&= \frac{iRB_0\sqrt{2\pi/3}-a}{2[j_1(Rk)-\tan\delta_1 n_1(Rk)]}\,,\label{c53}\\
-d_{10}\sin\delta_1&=\frac{iRB_0\sqrt{2\pi/3}-a}{2[-\cot\delta_1 j_1(Rk)+n_1(Rk)]}\,.\label{c54}
\gal
\end{subequations}
Substitution of equations \eq{c25}--\eq{c54}  into \eq{b25} and \eq{b35} furnishes the analytic expressions for the electromagnetic field of the spherical domain in the monochromatic uniform magnetic field. 

\subsection{Electric current and magnetic moment}\label{sec:d}

Using the results of the previous section one can compute the total current flowing in the direction of the external magnetic field through any cross sectional area of the domain:
\ball{d2}
I_\omega=  \sigma_\chi \int B_{\omega z}dS_z+ \sigma\int E_{\omega z}dS_z=\sigma_\chi \Phi_B + \sigma \Phi_E\,,
\gal
 The magnetic field flux  can be written as
\bal
\Phi_B = &2\pi \sigma_\chi \int_0^{\sqrt{R^2-z^2}} B_{\omega z} \rho d\rho =  2\pi \sigma_\chi \int_0^{\sqrt{R^2-z^2}} (\cos\theta B_{\omega r}-\sin\theta B_{\omega \theta}) \rho d\rho\nonumber\\
=&2\pi \sigma_\chi \int_z^R \left(\frac{z}{r} B_{\omega r}-\frac{\rho}{r} B_{\omega \theta}\right) r dr\,, \label{d4}
\gal
where $\rho$ is the radial coordinate in the cross-sectional plane and in the second line the integration variable has been changed  to $r= \sqrt{\rho^2+z^2}$. Using \eq{b25} and \eq{b19} one derives 
\bal
\Phi_B = &2\pi \left\{ z^2i\sqrt{\frac{3}{2\pi}}\int_z^R\frac{dr}{r^2}[g_{10}j_1(q_+r)+h_{10}j_1(q_-r)]\right.\nonumber\\
&-
\left.\frac{i}{2}\sqrt{\frac{3}{2\pi}}\int_z^R\frac{dr}{r^2}(r^2-z^2)\left[g_{10}(j_1(q_+r)r)'+h_{10}(j_1(q_-r)r)'\right]
\right\}\,.\label{d6}
\gal
Integrating the  second integral  by parts and using the boundary condition \eq{c34} yields
\ball{d8}
\Phi_B=0\,.
\gal
Thus, the anomalous component of the current does not contribute to the charge separation current. 

The computation of the electric flux can be done along the same lines by noting that $E_{\omega z}= i\omega A_{\omega z}$ and using \eq{b21} in place of \eq{b25}. The result is 
\ball{d10}
I_\omega= \sigma\Phi_E= -\sigma \omega \sqrt{\frac{3\pi}{2}}\frac{R^2-z^2}{R}j_1(Rq_+)(q_++q_-)g_{10}\,.
\gal
This constitutes the charge separation effect. The current $I_\omega$ does not identically vanish  as long as $\Theta\neq 0$, i.e.\ either $\Theta_0$ or $\sigma_\chi$ is finite. 

The magnetic moment of the domain is given by
\ball{d20}
\b \mu = \frac{1}{2}\sigma_\chi \int \b x\times \b B\, d^3x+\frac{1}{2}\sigma \int \b x\times \b E\, d^3x\,
\gal
and can be computed using the same steps as were employed in the calculation of the current. The result is
\ball{d22}
\b \mu_\omega  = &i\unit z\sqrt{\frac{2\pi}{3}}\left\{ \frac{g_{10}}{q_+^3}\left[ (3-R^2q_+^2)\sin(Rq_+)-3Rq_+\cos(Rq_+)\right]\left( \sigma_\chi + \frac{i\omega \sigma}{q_+}\right)\right.\nonumber\\
&\left. -\frac{h_{10}}{q_-^3}\left[ (3-R^2q_-^2)\sin(Rq_-)-3Rq_-\cos(Rq_-)\right]\left( \sigma_\chi - \frac{i\omega \sigma}{q_-}\right)
\right\}\,.
\gal
It vanishes if $\sigma_\chi \to 0$, i.e.\ existence of the domain magnetic moment requires the anomalous current.

\section{Application to heavy-ion collisions}\label{sec:e}

In this section we specialize the results of the previous section to the heavy-ion collisions phenomenology.  
The quark-gluon plasma produced in heavy-ion collisions is subject to external magnetic field induced by the spectator valence charges \cite{Kharzeev:2007jp,Skokov:2009qp,Voronyuk:2011jd,Ou:2011fm,Bzdak:2011yy,Bloczynski:2012en,Deng:2012pc,Tuchin:2010vs,Tuchin:2015oka,Stewart:2017zsu,Peroutka:2017esw}. The time-dependence of this field is quite complicated. It is convenient to adopt a simple parameterization introduced in \cite{Yee:2013cya,Yin:2015fca}
\ball{d3}
\b B^\text{ext}(t)= \frac{B_0\unit z}{1+(t/t_0)^2}= \frac{1}{2}B_0 t_0 \unit z \int_{-\infty}^\infty d \omega\, e^{-t_0|\omega|-i\omega t}\,,
\gal
where $t_0=0.6$~fm. It accounts for fact that an electrically conducting medium slows down the decay of the electromagnetic field  \cite{Tuchin:2010vs,Tuchin:2013ie,Tuchin:2013apa,Zakharov:2014dia,Tuchin:2015oka}. Magnetic field inside the domain follows from \eq{b25}
\ball{d5}
\b B^\text{in}(\b x,t)=  
\frac{1}{2} t_0 \unit z \int_{-\infty}^{+\infty} d \omega \, e^{-t_0|\omega|-i\omega t}
\left[ g_{10} q_+\b W_{10}^+(\b x, q_+)-h_{10}q_-\b W_{10}^-(\b x, q_-) \right]\,.
\gal
General properties of the magnetic field time-dependence can be inferred from the analytical structure of its Fourier component. The amplitudes $g_{10}$ and $h_{10}$ have poles at $Rq_+=x_n$ and $Rq_-=x_n$ correspondingly, where $x_n$, $n=0,1,2\ldots$  are zeros of the spherical Bessel function $j_1(x)$. The first three zeros are $x_1=4.49$, $x_2=7.73$ and $x_3=10.90$. 
The characteristic external field frequency $\omega_0\sim 1/t_0=1.7\,\text{fm}^{-1}$ is much larger than $\sigma$ and $\sigma_\chi$, which implies that the poles of $B_\omega^\text{in}$ are situated at $\omega \approx x_n/R$. Depending on the domain radius $R$ the integral over $\omega$ may pick up contributions from one or more poles. If $R<x_1/\omega_0 = 2.6$~fm, which is the phenomenologically most relevant case,  the magnetic field inside the domain is  suppressed by the factor $e^{-t_0 x_1/R}$. The magnetic field of domains with sizes $2.6< R <  4.6$~fm have the non-suppressed contributions of the first zero, while contributions of other zeros is still exponentially suppressed etc.

\begin{figure}
\begin{minipage}{.5\textwidth}
  \includegraphics[width=0.9\textwidth]{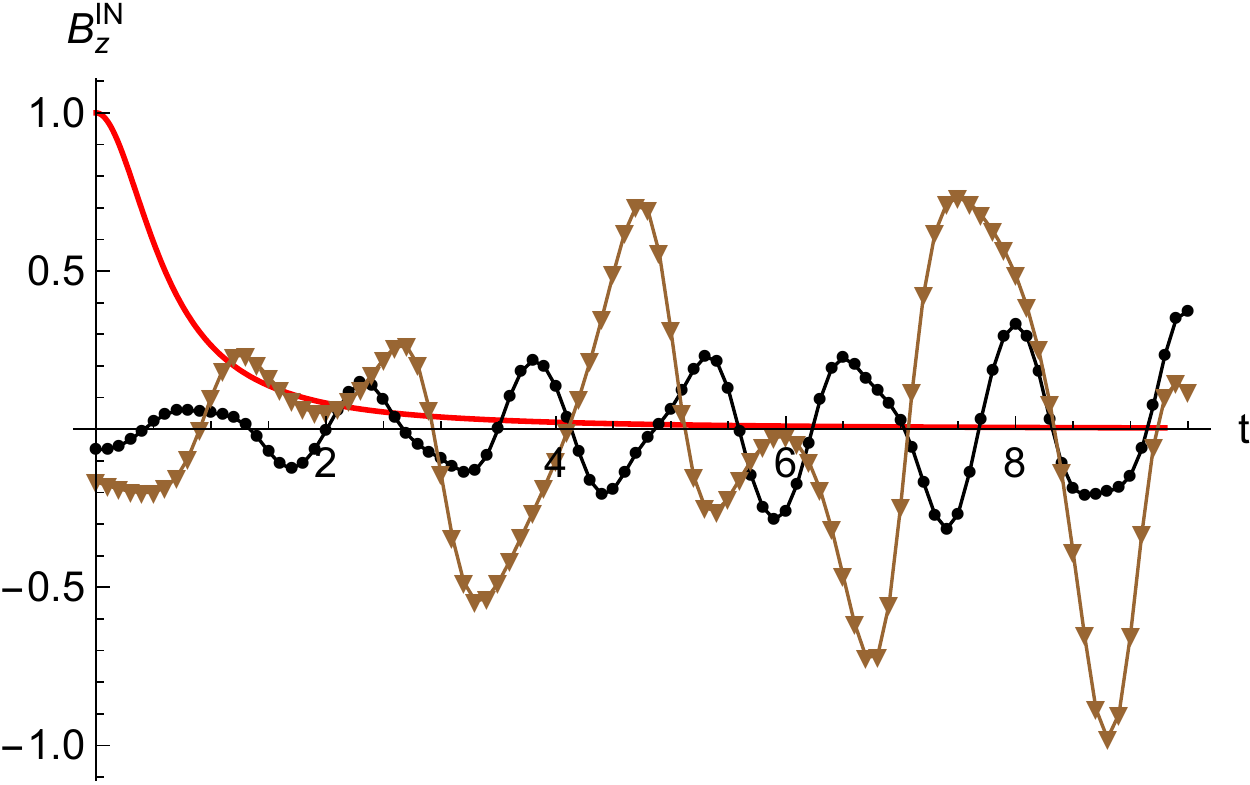}
\end{minipage}%
\begin{minipage}{.5\textwidth}
  \centering
  \includegraphics[width=0.9\textwidth]{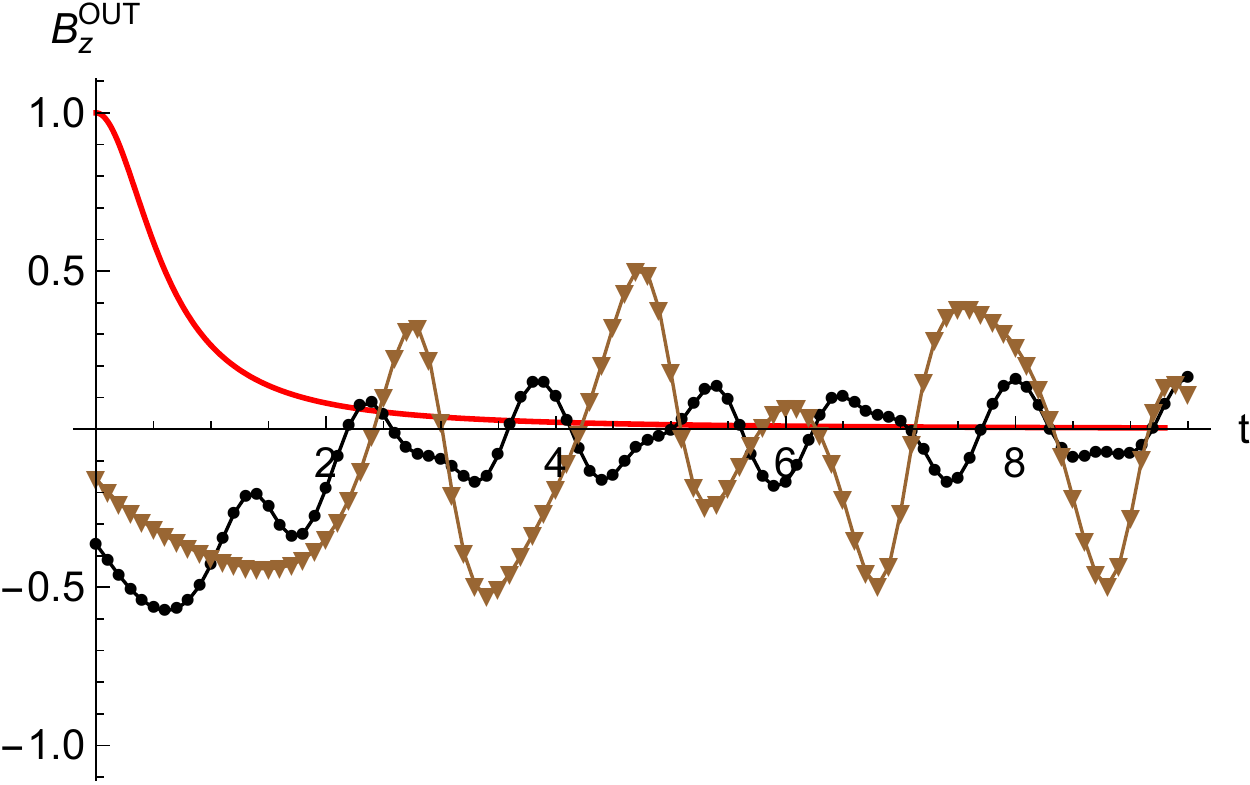}
\end{minipage}
\caption{\raggedright Left panel: $B_z$ inside a spherical domain at a representative point $r=R/2, \theta=\pi/3$.  Right panel: $B_z$ outside the domain at a representative point $r=3R/2, \theta=\pi/3$. Domain radius is $R=1$~fm (black dots) or $R=2$~fm (brown triangles). Other parameters: $B_0=1\,\text{fm}^{-2}$, $\sigma=1/(36\,\text{fm})$ \cite{Aarts:2007wj,Ding:2010ga,Amato:2013oja,Cassing:2013iz,Yin:2013kya}, $\sigma_\chi=1/(100\,\mathrm{fm})$, $\Theta_0=2\pi$. Solid red line represents the external field $B^\text{ext}$ of \eq{d3}.   }
\label{B-vs-time}
\end{figure}

This analysis is corroborated by the numerical calculation shown in \fig{B-vs-time}. It is seen that the induced field strength increases with the domain radius. It is worth noticing that even though the initial field decays at about 2~fm, the induced field oscillates long after that time due to low electrical conductivity of QGP. Actually, the oscillation amplitude of the magnetic field inside the domain increase indicating instability. This instability is caused by the brunch  cut singularity along the imaginary axis in the expression for $B_\omega^\text{in}$:
$$
\frac{i}{2}\left(-\sigma-\sqrt{\sigma^2+\sigma_\chi^2}\right)\le \omega \le \frac{i}{2}\left(-\sigma+\sqrt{\sigma^2+\sigma_\chi^2}\right)\,.
$$ 
This instability has been a subject of intensive study in recent years \cite{Joyce:1997uy,Boyarsky:2011uy,Tashiro:2012mf,Kharzeev:2013ffa,Akamatsu:2013pjd,Khaidukov:2013sja,Kirilin:2013fqa,Dvornikov:2014uza,Avdoshkin:2014gpa,Tuchin:2014iua,Sigl:2015xva,Buividovich:2015jfa,Manuel:2015zpa,Pavlovic:2016gac,Yamamoto:2016xtu,Xia:2016any,Kirilin:2017tdh,Rogachevskii:2017uyc}. It is established that the growth  of this instability is governed by the chiral anomaly equation. The unstable modes 
transfer helicity from the medium to the field in a process known as the inverse cascade \cite{Biskamp,Boyarsky:2011uy}. Eventually, however, the helicity conservation puts a cap on the inverse cascade \cite{Kaplan:2016drz,Tuchin:2017vwb}. As explained in \sec{sec:a}, this interesting effect is not really phenomenologically relevant for heavy-ion collisions. In fact, \eq{a10} explicitly neglects any significant long-time evolution effects.

\begin{figure}
\begin{minipage}{.5\textwidth}
  \includegraphics[width=0.7\textwidth]{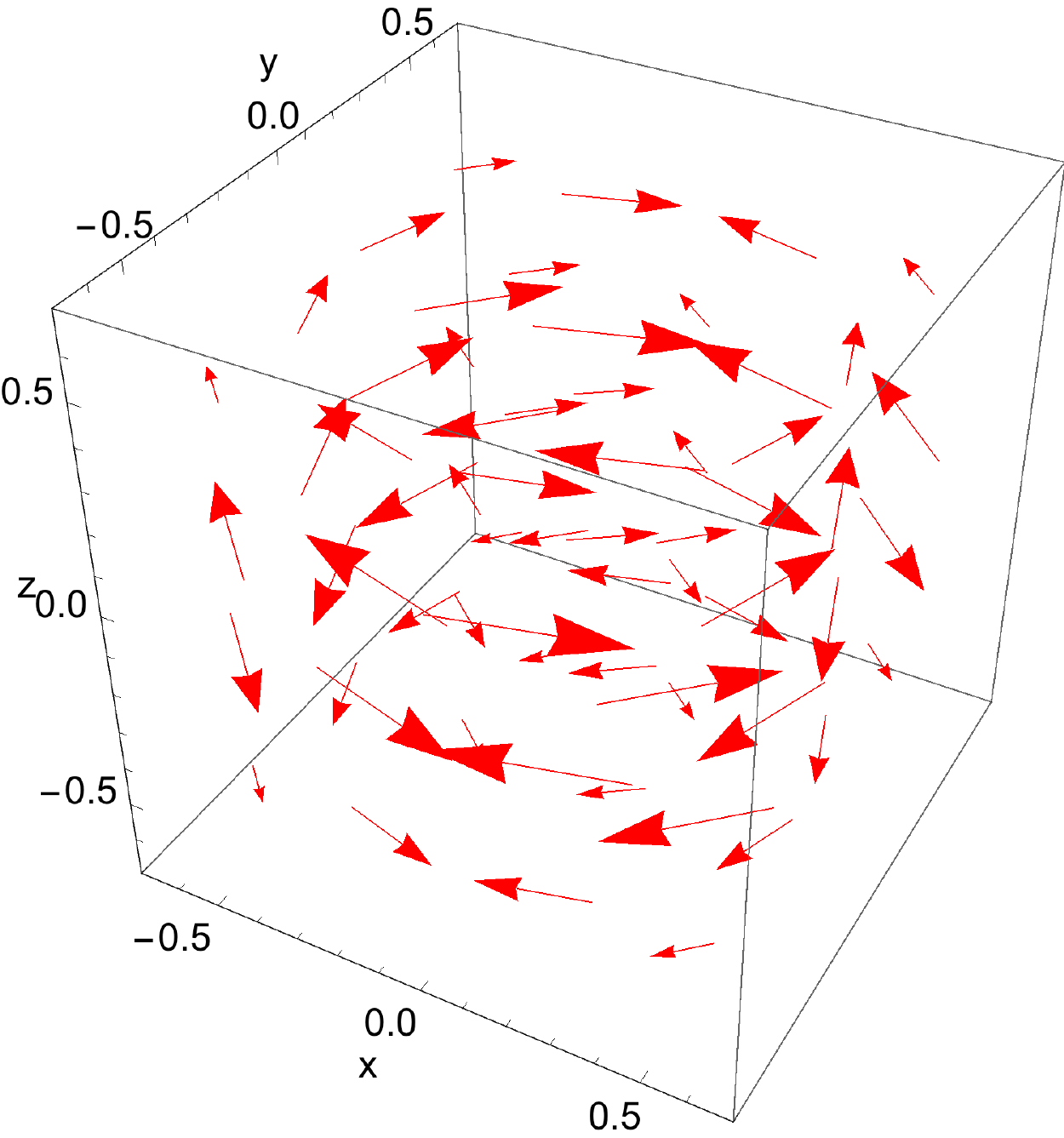}
\end{minipage}%
\begin{minipage}{.5\textwidth}
  \centering
  \includegraphics[width=0.7\textwidth]{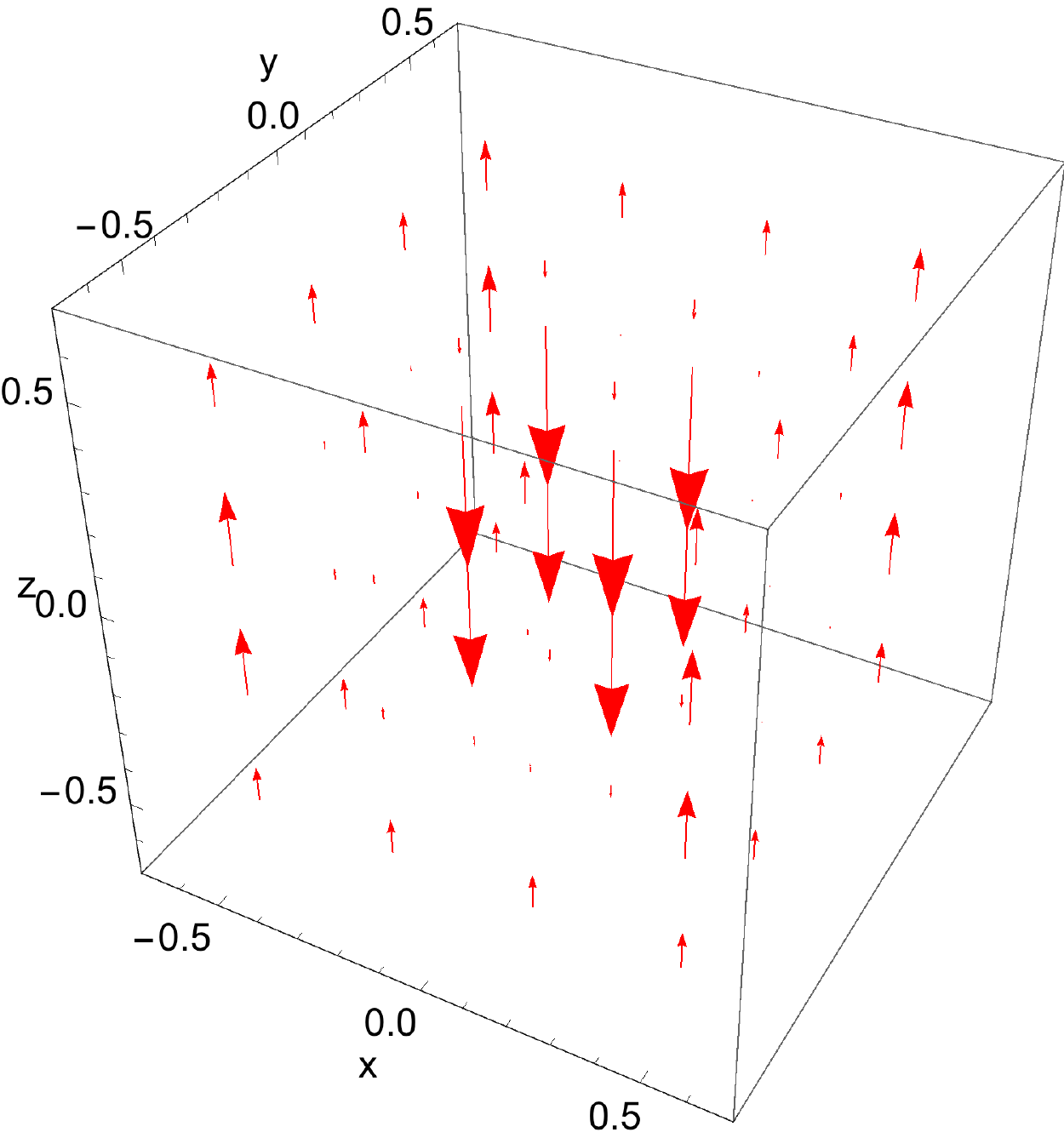}
\end{minipage}
\caption{\raggedright Snapshot of the magnetic field inside a spherical domain of radius $R=1$~fm at $t=2$~fm.  Left panel: $\b B$,  right panel: $B_z$ (zoomed in). Other parameters: $B_0=1\,\text{fm}^{-2}$, $\sigma=1/(36\,\text{fm})$, $\sigma_\chi=1/(100\,\mathrm{fm})$, $\Theta_0=2\pi$.  }
\label{3d-B}
\end{figure}

One can get a general idea about the magnetic field structure inside a spherical domain by looking at the snapshot shown in \fig{3d-B}. As can be expected, the field lines are mostly twisted around the direction of the external field owing to the smallness of the anomalous current. In order to better see the $z$-component of the magnetic field, the right panel magnifies it while discarding the transverse components. As has been shown in \sec{sec:c}, the magnetic field flux through the cross sectional area of the domain parallel to the $xy$ plane, vanishes. As the result, the number of magnetic field lines crossing in and out any $xy$ plane is equal.  This can be seen on the right panel as well. 

\begin{figure}
\begin{minipage}{.5\textwidth}
  \includegraphics[width=0.9\textwidth]{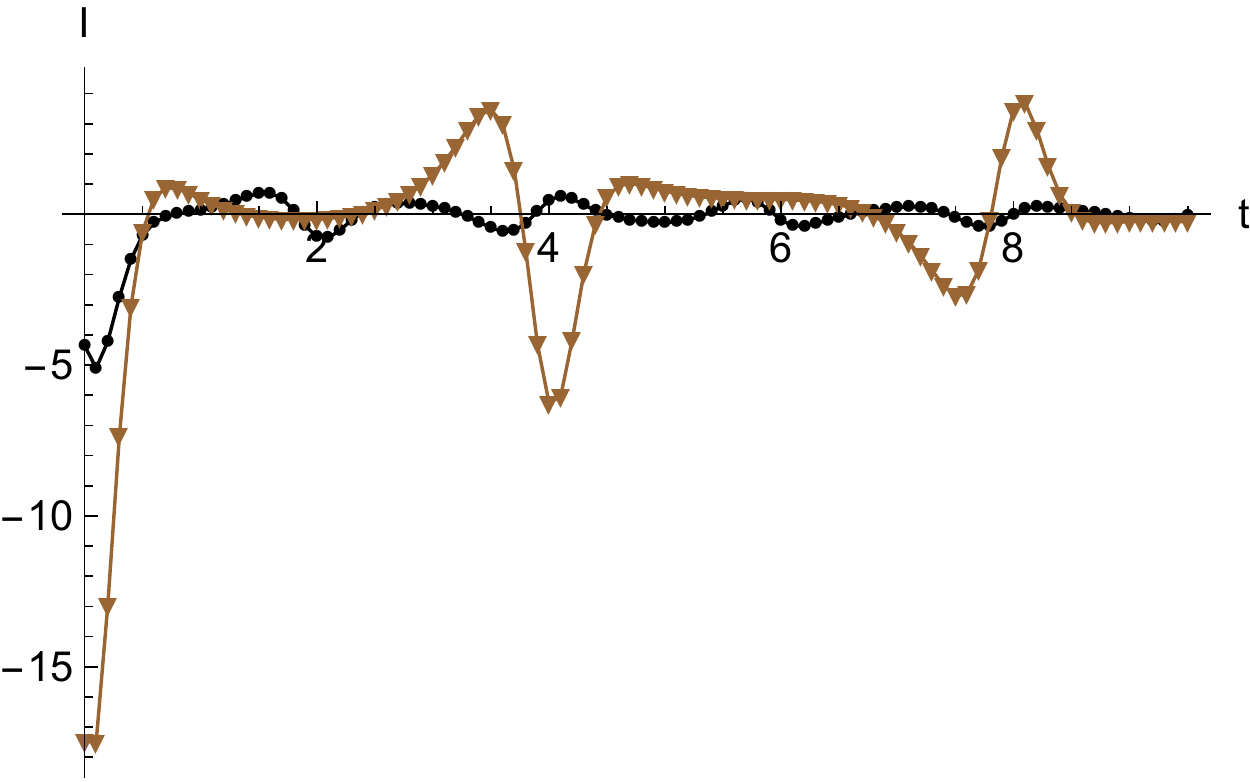}
\end{minipage}%
\begin{minipage}{.5\textwidth}
  \includegraphics[width=0.9\textwidth]{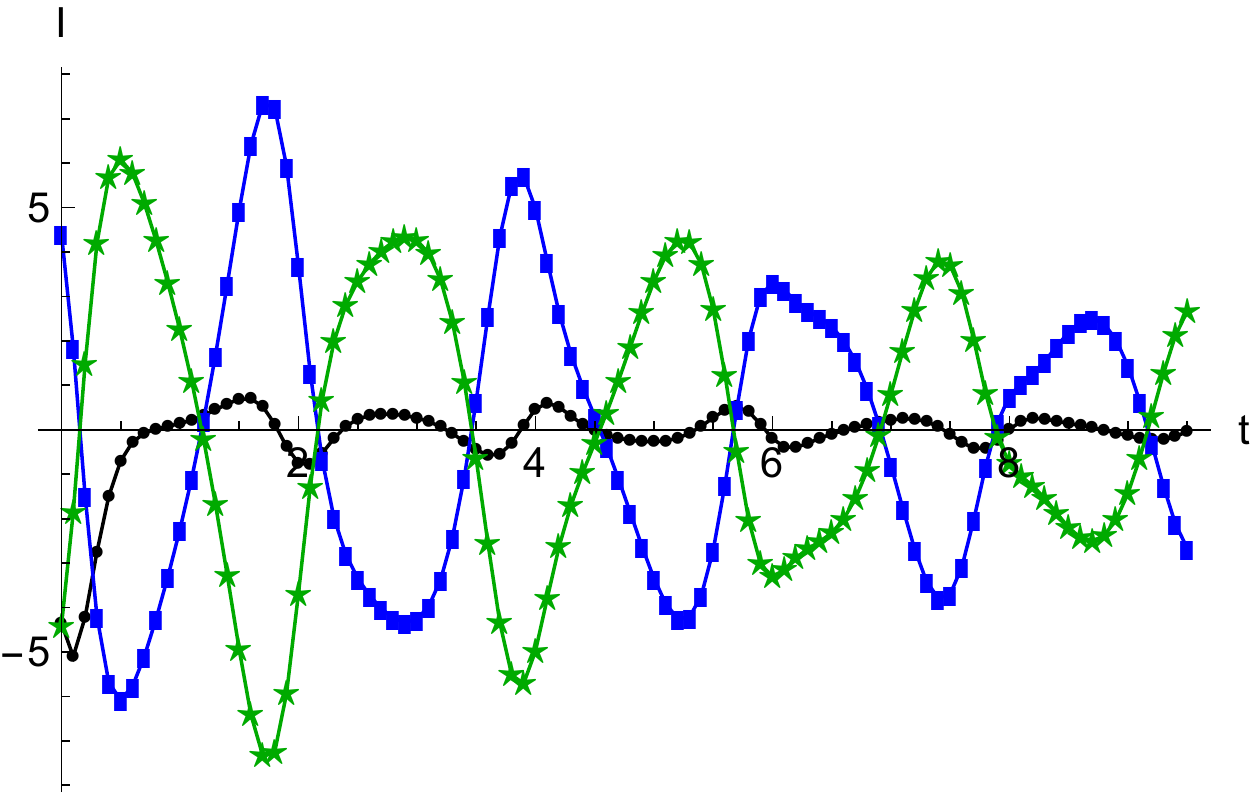}
\end{minipage}
\caption{\raggedright  Electric current flowing inside a domain in $z$-direction through a cross section at $z=R/2$.  $B_0=1\,\text{fm}^{-2}$, $\sigma=1/(36\,\text{fm})$. Black circles: $R=1$~fm, $\sigma_\chi=0.01/\mathrm{fm}$, $\Theta_0=2\pi$, brown circles: $R=2$~fm, $\sigma_\chi=0.01/\mathrm{fm}$, $\Theta_0=2\pi$, blue squares: $R=1$~fm, $\sigma_\chi=0.01/\mathrm{fm}$, $\Theta_0=-2\pi$, green stars: $R=1$~fm, $\sigma_\chi=-0.01/\mathrm{fm}$, $\Theta_0=2\pi$.  }
\label{current}
\end{figure}

Even though the magnetic field does not produce net electric current in the $z$-direction, the electric current does.  Induced electric current inside the domain is displayed in \fig{current} for a representative set of phenomenologically relevant parameters. One observes rapid oscillations of the current that may average to zero in a long run. Also, at any given time, an average value of the total current of a large enough ensemble of domains seems to average to zero. 

\section{Summary and discussion}\label{sec:s}

Metastable $CP$-odd topological domains emerge in the hot QCD matter.   The external magnetic field applied to these domains generates an anomalous current and charge densities. This paper focused on one such domain. To simplify the calculations, the domain was assumed to be a uniform sphere, while the surrounding medium to be  spatially uniform and topologically trivial. The electromagnetic field in  entire space was analytically calculated by employing a standard technique. The electric and magnetic components of the field induce Ohm and anomalous currents respectively. Their main properties are as follows. 

\textcolor{darkred}{1)} The normal component of the electric  current vanishes on the domain wall regardless of the domain geometry and uniformity. Thus no electric current flows into or out of the domain. 

\textcolor{darkred}{2)} The charge separation current, i.e.\ the total electric current flowing in the direction of the external magnetic field through any cross sectional area of the domain is the Ohm current, as shown in \sec{sec:d}. The contribution of the total anomalous current is zero. In particular, the total current vanishes in an electric insulator $\sigma\to 0$. This may appear counterintuitive because a $CP$-odd effect cannot be generated by the $CP$-even current. There is no contradiction though, as the the total current vanishes when $\Theta\to 0$. Even so, it is interesting to note that the current is finite if either $\Theta_0$ or $\sigma_\chi$ is finite. This is especially important if  $\sigma_\chi$ turns out to be much smaller than a few MeV as assumed in most applications; in that case the CME is generated by the domain walls.   

\textcolor{darkred}{3)} The total current is finite long after the external field decayed, owing to the low electrical conductivity of QGP, which implies small dissipation. The current oscillates with roughly the characteristic time $t_0$ of the external field. However, since no charge leaves the domain, the final charge separation within the domain depends on the current magnitude and direction at the time of the freeze-out. 

\textcolor{darkred}{4)} The resonance frequencies of a spherical domain are $\omega_n=x_n/R$, where $x_n$ are zeros of the spherical Bessel function $j_1(x)$. The current frequency modes with $\omega\ll \omega_1$ do not contribute to the total current as the corresponding wavelength does not fit in the domain. In the static limit $I_\omega\to 0$ as $\omega\to 0$.\footnote{Actually, the MCS equations do have non-trivial solutions even in the absence of the external field. These are given by the CK states \eq{b8}--\eq{b11} with $\alpha=\sigma_\chi$. However, their radii  $R_n=x_n/\sigma_\chi$ are way too big to fit into the QGP, as was first pointed out in \cite{Chernodub:2010ye}.}


Finally, the author believes that the present model, despite its simplicity,  gives a reasonably accurate idea about a possible effect of the domain size on the charge separation effect.  It has been seen throughout the paper that the properties enumerated above a fairly geometry independent. The gradients $\b\nabla \Theta$ also seem to be a minor  effect  \cite{Tuchin:2016qww}. It thus appears that giving up the spherical symmetry and spatial uniformity would not have a large impact on the above conclusions.

\acknowledgments
  This work was   was supported in part by the U.S. Department of Energy under Grant No.\ DE-FG02-87ER40371.



\end{document}